\documentstyle{article}
\begin{document}

\title{Structure and convergence of Poincar\'{e}-like normal forms.}
\author{S.Louies\thanks{Fax.: 32.2.650.58.24, phone: 32.2.650.59.03,
e-mail: slouies@ulb.ac.be}
 and L.Brenig\thanks{Fax.: 32.2.650.58.24, phone: 32.2.650.58.07,
e-mail: lbrenig@ulb.ac.be}.\\   \\  {\em Universit\'{e} Libre de
Bruxelles.
 Service de physique statistique CP231.} \\ {\em Campus de la Plaine.
 1050 Brussels-Belgium.}}
\date{ }
\maketitle
\begin{abstract} The general term of the Poincar\'{e} normalizing
series is explicitly constructed for non-resonant  systems of ODE's
in a large class of equations. In the resonant case, a non-local
transformation is found, which exactly  linearizes the ODE's and
whose series expansion always converges in a finite domain. Examples
are treated.\newline PACS code: 47.20.Ky.  \newline
\end{abstract}
\vspace{1cm}
\section{Normal forms and Poincar\'{e} transformation.} Let us
consider a system of nonlinear ordinary differential equations:
\begin{eqnarray}
\dot{x}_{i} = \lambda_{i}x_{i}+f_{i}(x),  \hspace{1cm} (i=1,...,N) 
\label{debut}
\end{eqnarray} where the (purely nonlinear) functions $f_{i}(x)$ are
analytic in the real variables $x_{1}...x_{N}$. We suppose, for the
simplicity of the purpose, that the linear part has been diagonalised.

The study of the qualitative behavior of the system starts with a
linear stability analysis of the fixed points \cite{Guckenheimer}.
This  determines the values of the control parameters for which the
system bifurcates (i.e. for which some stable fixed point becomes
unstable, or vice-versa.). A further  analysis gives the nonlinear
behavior of the solutions in the neighborhood of the fixed points.
Among the methods to compute this last step, the most often used  is
the normal form analysis \cite{Arnold}, that we shortly describe
here: consider a particular form of the system (\ref{debut}) for
which the origin is a fixed point; this implies  the following form:
\begin{eqnarray}
\dot{x}_{i}= \lambda_{i} x_{i} + \sum_{m}a_{i}(m) x^{m}, 
\hspace{1cm} (i=1,...,N)  \label{nf1}
\end{eqnarray} where we used the multi-index notation, $m \equiv
(m_{1},...m_{N})$, $a_{i}(m) \equiv a_{i}(m_{1},...,m_{N})$ and
$x^{m}\equiv x_{1}^{m_{1}}...x_{N}^{m_{N}}$. The multiple sum is
taken over the integers $m_{1},...,m_{N}$,  such that
$\sum_{j=1}^{N}m_{j}=\mid m \mid \geq 2$, with $m_{i}\geq 0$.
Consider now the near identity change of variables given by the
following formal series:
\begin{eqnarray} x_{i}=y_{i} + \sum_{m, \mid m \mid \geq 2} b_{i}(m)
y^{m} \label{nf2}
\end{eqnarray} This gives rise to a new system for the $y_{i}$:
\begin{eqnarray}
\dot{y}_{i}= \lambda_{i} y_{i} + \sum_{m, \mid m \mid \geq 2}c_{i}(m)
y^{m} ,  \hspace{1cm} (i=1,...,N)  \label{nf3}
\end{eqnarray} Now, let us choose the $b_{i}(m)$ in such a way that
this last system is as simple as possible. The ideal case would be
$c_{i}(m)=0$ for all $i$ and all $m$, but it is  easy to see that
this will be possible only if no {\it resonance condition} is
satisfied, i.e. if there exist no set of $N$ positive integers
$r_{i}$ with 
$\sum_{j=1}^{N}r_{j}\geq 2$, such that:
\begin{eqnarray}
\sum_{j=1}^{N} r_{j}\lambda_{j}-\lambda_{i} = 0 \label{rescond}
\end{eqnarray} for some $i$. The reason is that, substituting
(\ref{nf2}) in (\ref{nf1}) and requiring $c_{i}(m)=0$, for all $i$
and all $m$, we determine $b_{i} (m)$  by a relation where the factor
of $b_{i} (m)$ is precisely $(\sum_{j=1}^{N}
m_{j}\lambda_{j}-\lambda_{i})$. If this quantity is equal to zero,
$b_{i}(m)$ remains undetermined, and the system (\ref{nf3}) will
still be nonlinear, with:
\begin{eqnarray} c_{i}(m) & \ne &  0 ,  \hspace{1cm} \mbox{if
$\sum_{j}m_{j} \lambda_{j} - \lambda_{i} =0$}  \nonumber \\
         & =   &  0 ,  \hspace{1cm} \mbox{otherwise}   
\label{nf3cond}
\end{eqnarray} Equations (\ref{nf3}) with conditions (\ref{nf3cond})
are called {\it normal form}. They form, in fact, a system equivalent
to (\ref{nf1}), where all the {\it  non-resonant  monomial} have been
eliminated. The transformation (\ref{nf2}) is then called {\it
Poincar\'{e} transformation}, and the monomials remaining in
(\ref{nf3}), {\it resonant monomials}.

In most cases, the general term of the series (\ref{nf2}) giving the
Poincar\'{e} transformation is not known, and, usually, it is
computed order by order. When no resonance condition is satisfied,
 the solutions of (\ref{nf3}) are simply exponentials of time, and
(\ref{nf2}) provides a formal series of exponentials which is
solution of (\ref{nf1}). Moreover,  the convergence of (\ref{nf2}) is
guaranteed in two cases \cite{Arnold}: if the real part of all the
$\lambda_{i}$ are all of the same sign (Poincar\'{e} domain). Or if
the $\lambda_{i}$ belong to the Siegel domain, i.e. zero lies within
the convex hull of $\lambda_{1},...,\lambda_{N}$ in the complex
$\lambda$-plane, along with:
\begin{eqnarray}
\mid \lambda_{i}-\sum_{j=1}^{N} \lambda_{j}m_{j} \mid \geq
\frac{c}{\mid m \mid ^{\nu}}  \nonumber
\end{eqnarray} for some $c>0$, $\nu \geq \frac{N-2}{2}$.

Nonetheless, there are domains of the complex $\lambda$-plane for
which the convergence of the Poincar\'{e} series is not guaranteed.

Even if the series  is convergent, still one problem remains: the
initial conditions of the new system (\ref{nf3}) have to be
calculated in terms of the initial conditions  of the old system
(\ref{nf1}) and of its parameters $\lambda_{i}$ and $a_{i}(m)$. To
investigate this, we have to invert the Poincar\'{e} series, which is
not an easy task. In practice, one inverts a truncated series. This
leads to an error on the new initial conditions which may be
excessively large and not controlled.

In the resonant case, equation (\ref{nf3}) is no longer linear, and
contains in fact an infinity of monomials (since if one resonance
condition is satisfied, so are an infinity of others). The use of the
Poincar\'{e} series to approximate the solution of (\ref{nf1}) is
even more difficult, and suffers from the absence  of general
convergence theorems.

 From a geometrical point of view, the Poincar\'{e} transformation
defines a diffeomorphic transformation to coordinates in the
phase-space, in which the solutions are, in the non-resonant case,
exponentials. The possible divergence in the non-resonant case, and
the  impossibility of eliminating the resonant monomials express the
fact that the transformation leading to such a coordinate system is
not always a diffeomorphism.

The main goal of this paper is first to derive the general term of
the Poincar\'{e} series for an $n$-dimensional Lotka-Volterra (L.V.)
system in the  non-resonant case. That is, we find an exact solution
to the recursion relation for the coefficient of the Poincar\'{e}
series.
 In a second step, we build a Poincar\'{e}-like transformation such
that:

1-The system for the new variables is always linear, no matter if
resonance conditions are satisfied.

2-The initial conditions for the new system are known (we show that
they may be chosen equal to the initial conditions for the original
system.).

3-The convergence of the series giving the transformation is always
guaranteed in a domain around the initial time $t_{0}$.

The price to pay for such nice properties is to be found in the fact
that this transformation is non-local as will be discussed  later: it
does not define a diffeomorphism. But still, it has many attractive
features ranging from explicit computation  of solutions to more
theoretical questions in nonlinear dynamics.

The restriction to the L.V. format may look rather restrictive. This
is not the case, since it is known that a large class of systems can
be transformed into  L.V. systems  \cite{Peschel} \cite{Léon}
\cite{plus} \cite{annibal}.  We show in the next section how this can
be done.

\section{The quasi-monomial formalism.} We restrict our attention to
nonlinear systems that can be written as:
\begin{eqnarray}
\dot{x}_{i}=x_{i}(\alpha_{i} +\sum_{j=1}^{M}A_{ij}
\prod_{k=1}^{N}x_{k}^{B_{jk}}) ,  \hspace{1cm} (i=1,...N)
\label{QMsys}
\end{eqnarray} No particular restriction is made neither on the
$A_{ij}$ nor on the $B_{jk}$: they may be real or complex numbers.
$M$ is the number of {\em quasi-monomials} 
$\prod_{k=1}^{N}x_{k}^{B_{jk}}$ appearing in (\ref{QMsys}). In what
follows, we suppose that $M$ is finite. The results proposed in this
work may be extended  to the case of an infinite number $M$ of
quasi-monomials, provided some convergence conditions on the $A_{ij}$
are satisfied. The class (\ref{QMsys}) is thus quite general: many of
the systems of interest in physics belong to this category. Systems
of the type (\ref{QMsys}) are called Quasi-Monomial (Q.M.) systems.

The transformation to the L.V. format is made by adding $M$ new
variables to the system (\ref{QMsys}). These are defined as follow:
\begin{eqnarray} x_{N+k}=\prod_{l=1}^{N}x_{l}^{B_{kl}} , 
\hspace{1cm} (k=1,...,M) \label{QMT}
\end{eqnarray} Taking the derivative with respect to time, we find
the $(N+M)$ dimensional system:
\begin{eqnarray}
\dot{x}_{i}=\lambda_{i}x_{i} + x_{i}\sum_{j=1}^{N+M} M_{ij}x_{j}
\label{LV}
\end{eqnarray} with:
\begin{eqnarray}
\lambda_{i} & = &\alpha_{i}                      ,  \hspace{2.85cm} 
\mbox{for $1\leq i \leq N$}  \label{deflambda} \\
             &  &\sum_{k=1}^{N}B_{(i-N)k}\alpha_{k}  ,  \hspace{1cm} 
\mbox{for $N+1 \leq i \leq N+M$} \nonumber
\end{eqnarray} and:
\begin{eqnarray} M_{ij} & = & 0     ,  \hspace{3.07cm}  \mbox{for  $1
\leq  i  \leq N+M$; $1 \leq j \leq N$}  \nonumber \\
       & & A_{i(j-N)} ,  \hspace{2.06cm}  \mbox{for $1 \leq i \leq
N$; $N+1 \leq j \leq N+M$}  \label{defm} \\
       & & \sum_{k=1}^{N}B_{(i-N)k}A_{k(j-N)} ,  \hspace{0.2cm} 
\mbox{for $N+1 \leq i \leq N+M$; $N+1 \leq j \leq N+M$} \nonumber
\end{eqnarray} The system (\ref{LV}) is of the form of the well known
Lotka-Volterra equations, first introduced in theoretical ecology
\cite{May}.

Note that the linear term can be eliminated by adding one more
variable which is put equal to one ($\dot{x}_{N+M+1}=0$;
$x_{N+M+1}(t_{0})=1$). We then have:
\begin{eqnarray}
\dot{x}_{i}=x_{i} \sum_{j=1}^{N+M+1} \tilde{M}_{ij}x_{j} \label{LVbis}
\end{eqnarray} where:
\begin{eqnarray}
\tilde{M}_{ij}& = & M_{ij} ,  \hspace{0.73cm} \mbox{for $1 \leq i$,
$j \leq M+N$}  \nonumber \\
               &  & \lambda_{i},  \hspace{1cm} \mbox{for $1 \leq i
\leq M+N$; $j=M+N+1$} \nonumber \\ & & 0 ,  \hspace{1.16cm} \mbox{for
$i=M+N+1$; $1 \leq j \leq M+N+1$} \nonumber
\end{eqnarray} The main characteristic of the systems (\ref{LV}) and
(\ref{LVbis}) is that they present the lowest nonlinearity, i.e. the
quadratic one. Moreover, the quadratic term is not the most general
one (which would be of the form $\sum_{j,k}N_{ijk}x^{j}x^{k}$) but
still permits complex behavior like chaos.

It has been shown \cite{Léon} that the general structure for the
Taylor series coefficient can be obtained for the solution of
(\ref{LVbis}). One finds:
\begin{eqnarray}
x_{i}(t)&=&\sum_{n=0}^{\infty}c_{i}(n)\frac{(t-t_{0})^{n}}{n!} 
\nonumber \\ c_{i}(0)&=&x_{i}(t=t_{0}) =x_{i0} \nonumber \\ 
c_{i}(n)&=&
x_{i0}\sum_{i_{1}...i_{n}}\tilde{M}_{ii_{1}}(\tilde{M}_{ii_{2}}+\tilde{M}_{i_{1}i_{2}})...(\tilde{M}_{ii_{n}}  
         +\tilde{M}_{i_{1}i_{n}}+ ... \nonumber \\
        & &
+\tilde{M}_{i_{n-1}i_{n}})x_{i_{1}0}x_{i_{2}0}...x_{i_{n}0} \nonumber
\end{eqnarray} We will show, in the next section, that the general
structure of the Poincar\'{e} series coefficient is very close to
this one. In the last part of this paper, we also show that the
Taylor series is a particular case of a class of series expansions of
the solution which includes also the Poincar\'{e} series.

\section{Structural analysis of the Poincar\'{e} transformation.} We
take as starting point (since the linear part plays an important role
in the normal form analysis) the system:
\begin{eqnarray}
\dot{x}_{i}=\lambda_{i}x_{i} + x_{i} \sum_{i=1}^{N} M_{ij}x_{j} , 
\hspace{1cm} (i=1,...,N) \label{LVN}
\end{eqnarray} Where, from now on, $N$ stands for $N+M$ in the
previous expressions.

It is known, since the work of Carleman \cite{Carleman}
\cite{Leon-C}, that a nonlinear system can be viewed as an
infinite-dimensional linear system. This can be realized  by
considering as new variables all the monomials one can build with
products of positive integer powers of the $x_{i}$. Using the
multi-index notation:
\begin{eqnarray} X_{m}\equiv
x^{m}=x_{1}^{m_{1}}x_{2}^{m_{2}}...x_{N}^{m_{N}}  \label{Cvar}
\end{eqnarray} and taking the derivative of (\ref{Cvar}), one finds:
\begin{eqnarray}
\dot{X}_{m}=
(\sum_{k=1}^{N}m_{k}\lambda_{k})X_{m}+\sum_{p=1}^{N}(\sum_{l=1}^{N}m_{l}M_{lp})X_{m+e_{p}}
\label{Csys}
\end{eqnarray} where $e_{p}$ is a unit vector such that
$(e_{p})_{s}=\delta_{p,s}$ (i.e. $X_{m+e_{p}}=
X_{m_{1},...,m_{p}+1,...,m_{N}}$). The infinite-dimensional linear
system (\ref{Csys}) is characterized by a triangular matrix $R$ which
is given by:
\begin{eqnarray} R_{mp} = \sum_{o=1}^{N}m_{o}\lambda_{o}
\delta_{m,p}+\sum_{k=1}^{N}\sum_{l=1}^{N}m_{k}M_{kl}\delta_{m+e_{l},p}
\nonumber
\end{eqnarray}

 For an original system (\ref{LVN})  that would be linear (that is,
the matrix $M$ vanishes) the system (\ref{Csys}) would be purely
diagonal. This implies that the Poincar\'{e} transformation on
(\ref{LVN}), for non vanishing matrix $M$, but in absence of any
resonance, corresponds to the diagonalisation of the
infinite-dimensional matrix defined by the system (\ref{Csys}).

Consider now the relation:
\begin{eqnarray} L_{mp}=\delta_{m,p} + \frac{ \sum_{k} R_{mk}(1 -
\delta_{mk}) L_{kp} } { R_{pp} - R_{mm}+\delta_{m,p} } \label{PTC}
\end{eqnarray} where, once again, the indices are multiple (the sum
over $k$ is a multi-sum over the $N$ indices $k_{1},...,k_{N}$
running from $0$ to $\infty$, and 
$\delta_{m,p}$ stands for
$\delta_{m_{1},p_{1}}...\delta_{m_{N},p_{N}}$).

We claim that since $R_{mp}$ is triangular (that is: $R_{mp}=0$ if
there exists at least one integer $k$ between $1$ and $N$, such that
$m_{k}>p_{k}$) and if $R_{mm} \neq R_{pp}$ for all $m$,$p$ with
$m\neq p$, then $(L_{mp})$ diagonalises $(R_{mp})$.

The condition $R_{mm}\neq R_{pp}$ for $m\neq p$ implies that the
relation $$\sum_{k=1}^{N}\lambda_{k}r_{k} = 0$$  is satisfied only
if  $r_{k}=0$ for all $k$ and restricts thus the system (\ref{LVN})
to the non-case. That $(L_{mp})$ diagonalises $(R_{mp})$ means here
that, considering the inverse operator $L^{-1}_{mp}$ defined by:
\begin{eqnarray}
\sum_{k}L^{-1}_{mk}L_{kp} = \sum_{k}L_{mk}L^{-1}_{kp} = \delta_{mp},
\label{invrel}
\end{eqnarray} we have:
\begin{eqnarray}
\sum_{k,o} L^{-1}_{mk} R_{ko} L_{op} = R_{mp} \delta_{mp}
\label{diagrel}
\end{eqnarray}

The question of the existence of the inverse operator $L^{-1}$ is
obvious. This operator represents the inverse of the Poincar\'{e}
transformation. The latter is a diffeomorphism. Hence, when it
exists, so does its inverse.

Let us now prove this proposition:

Multiplying both sides of (\ref{PTC}) by $(R_{pp} -
R_{mm}+\delta_{m,p})$ one finds:
\begin{eqnarray}
\sum_{k}R_{mk}L_{kp}  =  L_{mp}R_{pp} + \delta_{mp} (L_{mp} - 1)    -
\delta_{mp} (R_{pp}-R_{mm}) \label{step}
\end{eqnarray} We will now show that the last two terms of
(\ref{step}) are equal to zero. It is clearly the case for
$\delta_{mp} (R_{pp}-R_{mm})$. For $\delta_{mp} (L_{mp} - 1)$ we now
demonstrate that, thanks to the fact that $R_{mp}$ is triangular, so
is also $L_{mp}$, and all the elements $L_{mm}$ are equal to one. To
show this, consider the series obtained by iterating (\ref{PTC}).
This will be a series of powers of $ R_{mp}(1 - \delta_{mp})$ (with
coefficients depending on $m$ and $p$).  Now, the elements of the
$k^{th}$ power of $ R_{mp}(1 - \delta_{mp})$ for which $p \leq m+k$
are all equal to zero. This implies that the only contribution to 
$L_{mm}$ comes from the first term of the series, that is
$\delta_{mp}$. So, $L_{mm}=1$, and $\delta_{mp} (L_{mp}-1)=0$.
Multiplying then (\ref{step}) by
$L^{-1}_{om}$  and summing over $m$, we find the announced result.

Taking this result back to the original L.V. system, we can build the
Poincar\'{e} series. We insert in (\ref{PTC}), the matrix $R$ defined
by (\ref{Csys}):

and, writing (\ref{PTC}) for $m=e_{i}$, we find after some simple
algebra:
\begin{eqnarray} x_{i}&=&y_{i}\sum_{n=0}^{\infty}
\sum_{i_{1}=1}^{N}\sum_{i_{2}=1}^{N}...\sum_{i_{n}=1}^{N}
[M_{ii_{1}}(M_{ii_{2}}+M_{i_{1}i_{2}}) 
...(M_{ii_{n}}+M_{i_{1}i_{n}}+ ... \nonumber \\ & &
+M_{i_{n-1}i_{n}})] 
[\lambda_{i_{n}}(\lambda_{i_{n-1}}+\lambda_{i_{n}})...(\lambda_{i_{1}}+
...+\lambda_{i_{n}})]^{-1}  y_{i_{1}}y_{i_{2}}...y_{i_{n}} \label{PT}
\end{eqnarray} where the term corresponding to $n=0$ is, by
convention, set equal to one. 

This result can be readily extended to the system (\ref{QMsys}), by
using  the definitions  (\ref{deflambda}) and (\ref{defm}). If the
system (\ref{QMsys}) is polynomial, the obtained series is the
Poincar\'{e} series. But the class of systems given by (\ref{QMsys})
also contains non-polynomial systems. In these cases, the series
which is derived is more general than the Poincar\'{e} transformation
of (\ref{QMsys}).

Comparing this to the Taylor series, we see that the two structures
are very close. This is not surprising: the Poincar\'{e} series is,
in fact, the Taylor series  in which an infinity of resummations have
been performed. 

Accordingly with what we said about the Poincar\'{e} series,
(\ref{PT}) is not always convergent, as we shall see in the examples
treated later. Moreover, this  result is not valid in the resonant
case, but in some sense, can be extended to it, as we now see.

\section{The Poincar\'{e} transformation revisited.}

Let us define the $N$ variables:
\begin{eqnarray} u_{i}(t)=x_{i}(t)e^{- \int_{\gamma}^{t}d\beta
\sum_{j=1}^{N}M_{ij}x_{j}(\beta)} \hspace{0.5cm} (i=1,...N)
\label{MPT}
\end{eqnarray} where the $x_{i}$ satisfies (\ref{LVN}). Taking the
derivative of (\ref{MPT}) with respect to time, we see that the
differential
 system for $u_{i}$ is always linear, 
$\dot{u}_{i}=\lambda_{i}u_{i}$, no matter if resonance conditions are
satisfied, and independently of the value of $\gamma$.  As announced,
the definition (\ref{MPT}) is non-local. This is expressed by the
presence of the integral: the value of 
$u_{i}(t)$ depends on the values of $x_{i}(t')$ for $t'$ between
$\gamma$ and $t$. The inverse transformation (\ref{twenty})  reveals
this nonlocality in an even clearer way.

Choosing $\gamma$ equal to the initial time $t_{0}$, 
$u_{i}(t)$ and $x_{i}(t)$ have the same initial conditions. Moreover,
(\ref{MPT}) is a recursion relation defining  $x_{i}(t)$ in term of
$u_{i}(t)$. It is possible to prove that (\ref{MPT}) generates the
series:
\begin{eqnarray} x_{i}&=&u_{i}(t)\sum_{n=0}^{\infty}
\sum_{i_{1}=1}^{N}\sum_{i_{2}=1}^{N}...\sum_{i_{n}=1}^{N}
M_{ii_{1}}(M_{ii_{2}}+M_{i_{1}i_{2}}) ...(M_{ii_{n}}+M_{i_{1}i_{n}}+
...  \nonumber \\ & &
+M_{i_{n-1}i_{n}})\int_{t_{0}}^{t}d\alpha_{1}\int_{t_{0}}^{\alpha_{1}}d\alpha_{2}...\int_{t_{0}}^{\alpha_{n-1}}d\alpha_{n}
\nonumber \\ & &
[u_{i_{1}}(\alpha_{1})u_{i_{2}}(\alpha_{2})...u_{i_{n}}(\alpha_{n})]
\label{twenty}
\end{eqnarray} Here again, the explicit expressions of (\ref{MPT})
and (\ref{twenty}) for the system (\ref{QMsys}) are obtained using
(\ref{deflambda}) and (\ref{defm}). 

Comparing this to (\ref{PT}), we see that the $\lambda_{i}$ appearing
in the denominator, disappeared in the integrals of (\ref{twenty}).
Moreover these integrals are finite,  since the integrands are
exponentials, and the domains of integration are finite. Using the
explicit form of $u_{i}(t)$, we have:
\begin{eqnarray}
x_{i}&=&x_{i}(t_{0})e^{\lambda_{i}(t-t_{0})}\sum_{n=0}^{\infty}
\sum_{i_{1}=1}^{N}...\sum_{i_{n}=1}^{N}M_{ii_{1}}(M_{ii_{2}}+M_{i_{1}i_{2}})
\nonumber \\ & &...(M_{ii_{n}}+M_{i_{1}i_{n}}+ ...
+M_{i_{n-1}i_{n}})x_{i_{1}0}...x_{i_{n}0} \nonumber \\ & &
\int_{t_{0}}^{t}d\alpha_{1}\int_{t_{0}}^{\alpha_{1}}d\alpha_{2}...\int_{t_{0}}^{\alpha_{n-1}}d\alpha_{n}e^{\sum_{p=1}^{n}\lambda_{i_{p}}(\alpha_{p}-t_{0})}
\label{convPT}
\end{eqnarray} The integrals appearing in (\ref{convPT}) can be
performed in closed form by using Carlson special functions
\cite{Carlson}, but we do not need this result for our present
purpose. Moreover, it can be proven that this series always converges
on a finite domain around $t_{0}$, even in the resonant case. These
results will be detailed in a forthcoming publication \cite{prep}.

We can ask under which conditions on $\gamma$ the recursion
(\ref{MPT}) generates the series (\ref{PT}). To answer this question,
let us substitute  (\ref{PT}) in (\ref{MPT}). The integral can then
be performed order by order, and we see that $u_{i}(t)$ will be equal
to $y_{i}(t)$ if $\gamma$ is fixed  in such a way that the constant
term appearing after performing the integral (i.e. the primitive of
$\sum_{j=1}^{N}M_{ij}x_{j}$ evaluated at $t=\gamma$),  is equal to
zero. If the real parts of the $\lambda_{i}$'s are all positive
(resp. negative) we have to put $\gamma=-\infty$ (resp. $+\infty$),
since the constant
 term is a sum of exponentials of $\gamma$. But if there exist
$\lambda_{i}$'s whose real parts are of opposite sign, there will be
exponentials with positive  arguments, and others with negative
arguments. In this case, if $\gamma$ tends to infinity (no matter if
it is $+\infty$ or $-\infty$), some exponentials will  be divergent.
Identically, in the resonant case, some monomials built on the
variables $y_{i}$'s are constant, and no value of $\gamma$ will
fulfill the requirement.

In fact, the correspondence between (\ref{PT}) and (\ref{convPT}) is
more subtle. A bijective correspondence can be established between
the terms appearing
 in (\ref{PT}) and rooted tree graphs. In the graph formalism,
(\ref{MPT}) represents a relation between these graphs \cite{prep}.

An easy way to see why (\ref{PT}) can be divergent even in the
non-resonant case, and in the same blow, why (\ref{MPT}) is always
convergent in a certain domain, is the following: series (\ref{PT})
is a Taylor series expansion (in powers of $y_{i}$) of the
Poincar\'{e} transformation around the origin in the $y$-space.  But
nothing secures that the trajectory of the solution will go through
the origin. On the other hand, by definition, (\ref{MPT}) guarantees 
that the series reduces to its order zero at $t=t_{0}$. In other
words, the solution in the $y$-space is, at $t=t_{0}$, precisely at 
the point around which the series is built.

 Finally, as announced, when all the $\lambda_{i}$ tend to zero, the
integrals in (\ref{convPT}) generate the factors $(t-t_{0})^{n}/n!$,
and (\ref{convPT})  is precisely the Taylor series given above,
whereas series (\ref{PT}) for $\lambda_{i}$ tending to zero diverges.

Thus, although the transformation (\ref{MPT}) is non-local, its
structure explains many aspects related to the existence and
convergence  of the Poincar\'{e} transformations. Moreover, it leads
to useful algorithms for explicit calculation of solutions.

\section{Example: Ricatti projective systems.} As an example, we now
treat a particular case of the system (\ref{LVN}):  a class of
$N$-dimensional, integrable systems,  called projective Ricatti
systems \cite{Bountis} \cite{Reid}:
\begin{eqnarray}
\dot{x}_{i}= \lambda_{i} x_{i} + x_{i} \sum_{j=1}^{N}c_{j}x_{j}
\hspace{1cm} (i=1,...,N) \nonumber
\end{eqnarray} The usual Poincar\'{e} series (\ref{PT}) gives:
\begin{eqnarray} x_{i}(t) = y_{i}(t) \sum_{n=0}^{\infty}
(\sum_{j=1}^{N} \frac{c_{j}y_{j}(t)}{\lambda_{j}})^{n} \nonumber
\end{eqnarray} Thanks to the fact that the general solution of the
system is known, we can determine the initial conditions for the
$y_{i}$. This leads to:
\begin{eqnarray} x_{i}(t)  = 
\frac{x_{i0}}{1-\sum_{j=1}^{N}\frac{c_{j}x_{j0}}{\lambda_{j}}}
e^{\lambda_{i}(t-t_{0})} 
\sum_{n=0}^{\infty}(\frac{\sum_{j=1}^{N}\frac{c_{j}x_{j0}}{\lambda_{j}}e^{\lambda_{j}(t-t_{0})}}{1-\sum_{k=1}^{N}\frac{c_{k}x_{k0}}{\lambda_{k}}})^{n}
\nonumber
\end{eqnarray} It is clear that, for some choice of the $c_{i}$,
$\lambda_{i}$, and $x_{i0}$, this series will be divergent for any
$t$.  For example, in two dimensions, if we choose $\lambda_{1}$ real
and positive ($\lambda_{1}=l$), $\lambda_{2}=-\lambda_{1}$,
$x_{10}=x_{20}=s$, and $c_{1}=1=-c_{2}$,  the series becomes:
\begin{eqnarray} x_{i}(t) = \frac{s}{1-2s/ l}
e^{\lambda_{i}(t-t_{0})}\sum_{n=0}^{\infty}(\frac{s/l(e^{l(t-t_{0})}+e^{-l(t-t_{0})})}{1-2s/l})^{n}
\nonumber
\end{eqnarray} And the radius of convergence is given by the
inequality:
\begin{eqnarray}
\mid \frac{2}{l/s-2} ch(l(t-t_{0})) \mid < 1 \nonumber
\end{eqnarray} If $l>4s$, there exist no $t$ for which the above
relation is satisfied.

Consider now the series (\ref{convPT}): in this case, it gives:
\begin{eqnarray}
x_{i}(t)=x_{i0}e^{\lambda_{i}(t-t_{0})}\sum_{n=0}^{\infty}(\sum_{j=1}^{N}
\frac{c_{j}x_{j0}}{\lambda_{j}}(e^{\lambda_{j}(t-t_{0})}-1))^{n}
\nonumber
\end{eqnarray} which is always convergent on a domain
$]t_{0}-\tau,t_{0}+\tau[$. Moreover, if one of the $\lambda_{i}$'s
tends to zero, the limit can be taken without difficulty (and is
finite), which is not the case for the usual Poincar\'{e} series
given above.
\vspace{1cm}

\underline {Acknowledgment:}
\newline One of us (L.B.) beneficiated of the financial support of
the Euratom-Belgium Association on Fusion.

We thank A. Figueiredo from the Universidade National de Brazilia for
fruitful discussions.

\end{document}